\documentclass{desyproc}

\begin{document}
\title{Dish Antenna Searches for WISPy Dark Matter: Directional Resolution Small Mass Limitations}

\author{{\slshape Joerg Jaeckel$^1$, Stefan Knirck$^{1,2,*}$}\\[1ex]
$^1$Institute for Theoretical Physics, Heidelberg University, Germany\\
$^2$Department of Physics, Graduate School of Science, The University of Tokyo, Japan\\
$^*$speaker}

\contribID{knirck\_stefan}

\confID{13889}  
\desyproc{DESY-PROC-2016-XX}
\acronym{Patras 2016} 
\doi  

\maketitle

\begin{abstract}
	Hidden photon and axion-like dark matter may be detected using spherical reflective surfaces such as dish antenna setups converting some of the dark matter particles into photons and concentrating them on a detector. These setups may be used to perform directional searches measuring the dark matter momentum distribution. We briefly review the photon distribution one expects to detect with such an antenna and directional resolution in ray approximation. Furthermore we consider the regime $m_{DM} \lesssim (R_{sp}\,v_{DM})^{-1}$ where this approximation does not hold anymore due to the photon wavelength exceeding the expected distribution widths. We discuss how this affects the expected distributions and experimental implications.
\end{abstract}

\section{Introduction}

Hidden photons and Axion-like-particles (ALPs) turn out to be prime WISPy dark matter candidates, c.f. for example \cite{Nelson:2011sf,Arias:2012az}.
Hidden photons can be described by the following kinetic mixing Lagrangian~\cite{Holdom:1985ag} (for a review c.f., e.g.,~\cite{Jaeckel:2013ija})
\begin{equation}
{\mathcal{L}}=-\frac{1}{4}F^{\mu\nu}F_{\mu\nu}-\frac{1}{4}X^{\mu\nu}X_{\mu\nu}-\frac{\chi}{2}F^{\mu\nu}X_{\mu\nu}+m^{2}_{X}X^{\mu}X_{\mu}+j^{\mu}A_{\nu},
\end{equation}
where $F^{\mu\nu}$ and $X^{\mu\nu}$ are the ordinary and hidden photon field strength linked to the gauge fields $A^{\mu}$ and $X^{\mu}$, respectively. $m_{X}$ is the mass of the hidden photons and $j^{\mu}$ denotes the coupling to matter.
Solving the equations of motion one finds for a hidden photon momentum mode ${\mathbf k}_{DM}$ 
\begin{equation}
\label{eq:hpwave}
\left(\begin{array}{c}
{\mathbf E}
\\
{\mathbf E}_{\rm hid}
\end{array}\right)
={\mathbf E}_{\rm DM}\left(\begin{array}{c}
-1
\\
1/\chi
\end{array}
\right)\exp(-i(\omega t -{\mathbf k}_{DM}{\mathbf x}))
\end{equation}
$\mathbf{E}$ describes the ordinary electric field coupled to matter and ${\mathbf E}_{\rm hid}$ the hidden photon field. Analogous equations are obtained for the Axion in presence of a magnetic field.

Considering a plane conducting surface (or a mirror) the ordinary electromagnetic part in Eq.~\eqref{eq:hpwave} will cause electron movements on the surface, such that
\begin{equation}
\label{eq:boundary}
0={\mathbf E}_{\rm tot, \parallel}|_{\rm surface}.
\end{equation}

We showed in \cite{Jaeckel:2015kea} (c.f. also \cite{Jaeckel:2013sqa}) that for a dark matter incident angle $\alpha$ this causes an emission of a plane electromagnetic wave under the angle
\begin{equation}
\label{eq:inout}
\sin(\beta)=\sin(\alpha)\frac{v}{\sqrt{1+v^2}},
\end{equation}
with the dark matter velocity $v=\frac{|{\mathbf{k}}_{DM}|}{m_{X}}$. Therefore, the emission occurs almost perpendicular to the surface and can therefore be focused by a spherical surface in its center \cite{Horns:2012jf}.
Recent experiments using this method are for example described in \cite{Dobrich:2014kda,Suzuki:2015vka} and an advanced setup is discussed in~\cite{TheMADMAXWorkingGroup:2016hpc,Millar:2016cjp}.

For a spherical mirror with a curvature radius $R \gg \lambda_{DM} = \frac{2 \pi}{k_{DM}} = \frac{2 \pi}{v m_X}$ we can calculate the intensity distribution in the center of the sphere in the ray approximation and obtain for a dark matter incident with angle $\theta$ rotated around the y-axis of the detector and velocity $v$
\begin{equation}
\label{dist}
I(x,y)=\bigg\{\begin{array}{lcl}
\frac{1}{2 \pi}\frac{|\cos(\theta) v|}{\left(\left[x-R\sin(\theta)v\right]^2+y^2+R^2\cos^{2}(\theta)v^2\right)^{3/2}} & {\rm for}& 
\vartheta_{0}\leq \vartheta_{sp, max}
\\
0 & {\rm for} & \vartheta_{0}> \vartheta_{sp, max}
\end{array}
\end{equation}
~\begin{wrapfigure}[15]{r}[1em]{0.45\textwidth}
	\vspace{-1.5em}
	\centerline{\includegraphics[width=0.45\textwidth]{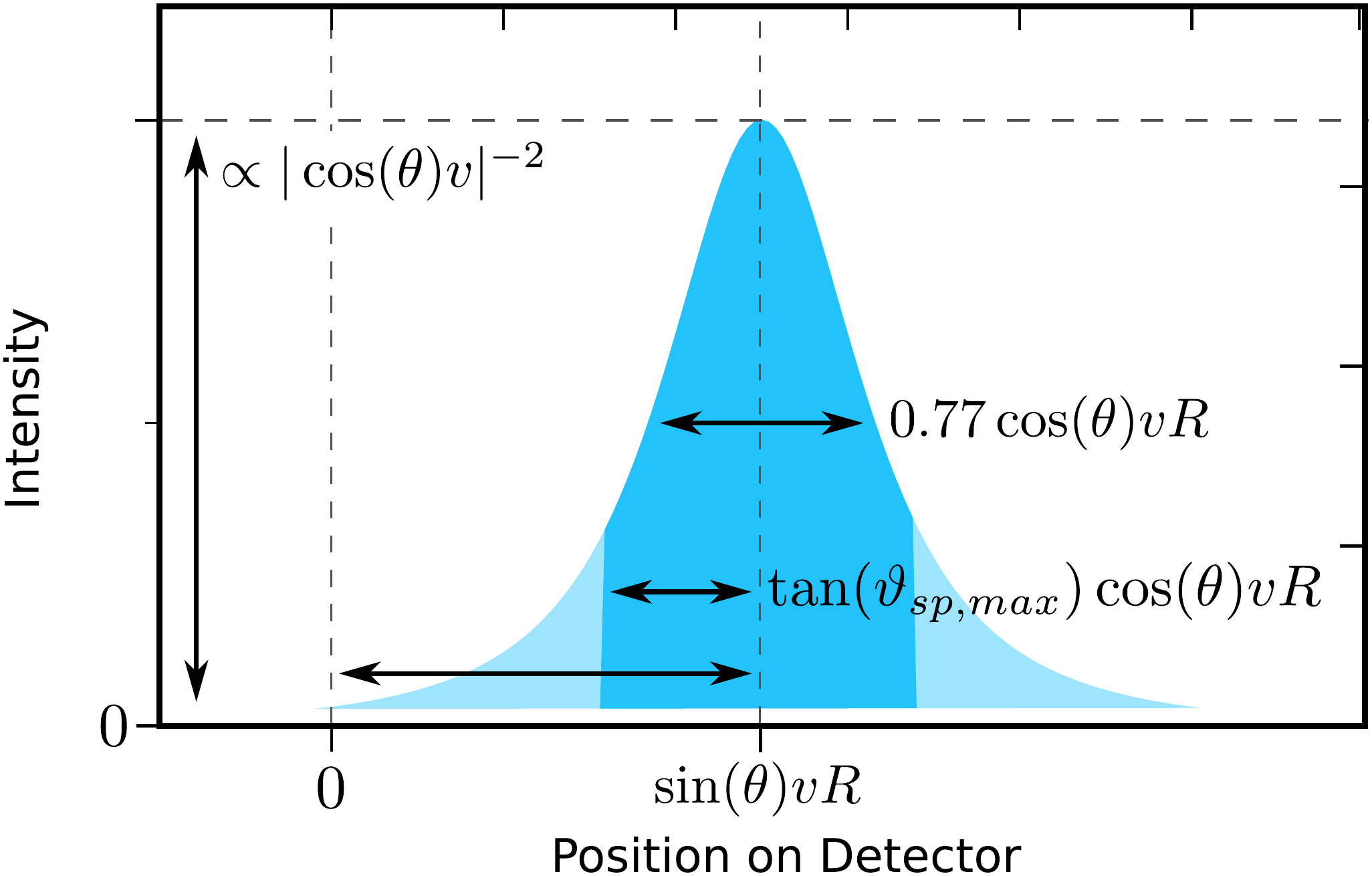}}
	\caption{Directional Resolution of dark matter particles with velocity $v$ coming from an angle $\theta$ to the detector normal in the ray approximation regime. For a spherical cap the tails are cut off so that the distribution is limited to the dark blue area.}
	\label{fig:directional-dist}
\end{wrapfigure}
where $\vartheta_{sp,max}$ is the aparture angle of the sphere; the result is normalized to the total power flux generated by a full half sphere with $\vartheta_{sp,max}=\pi/2$, and
\begin{equation}
\vartheta_{0}=\left|\arctan\left(\frac{\sqrt{(x-R\sin(\theta)v)^2 + y^2}}{R\cos(\theta)v}\right)\right|.
\end{equation}
Figure~\ref{fig:directional-dist} illustrates its basic properties. 

This may be used to perform a directionally sensitive dark matter search and gives direct implications on the directional resolution, but also on discovery experiments. For more details c.f. \cite{Jaeckel:2015kea}.
In this note we are going to investigate the validity of this result for long dark matter wavelengths $\lambda_{DM} \gtrsim R$ to infer limitations on the directional sensitivity. 

\section{Wave Ansatz}
For a correct calculation at high wavelengths one has to integrate over elementary waves emitted from the surface of the sphere.
Again, this ansatz needs to fulfill Eq.~\eqref{eq:boundary}.
One can show that for a plane surface the far-field part of a classical dipole field is sufficient when integrating over the whole surface.
At distance $\mathbf{r}$ from its origin it is given by (except time dependence)
\begin{equation}
\mathbf{E}_d(\mathbf{r};\mathbf{p}) = {{k_\gamma}^2} ( \hat{\mathbf{r}} \times \mathbf{p} ) \times \hat{\mathbf{r}}  ~ \frac{\exp({i {k_\gamma} r})}{r}  ~~~~;~~~~\mathbf{B}_d(\mathbf{r};\mathbf{p}) = {k_\gamma}^2 ~ \hat{\mathbf{r}} \times \mathbf{p} ~ \frac{\exp({i {k_\gamma} r})}{r} ,
\end{equation}
where $k_\gamma$ is the photon wavenumber and $\mathbf{p}$ the dipole-vector.
In order to fulfill~\eqref{eq:boundary} one needs to set (up to constant and spatial phase which we discuss below)
\begin{equation}
	\mathbf{p} ~ =   \frac{1}{2 \pi} \frac{\mathbf{E}_{\text{in,}||}}{{k_\gamma}}.
\end{equation}
Thus, $\mathbf{p}$ is a scaled projection of the dark matter polarization vector\footnote{We show simulations with circular transversal polarization as this behaves similar to arbitrary averaged polarization. When incoherently averaging over all possible polarizations the limitations for directional resolution are qualitatively the same. 
	} on the spherical surface.

The electric field at detector position $\mathbf{x}$ is then given by
\begin{equation}
\mathbf{E}(\mathbf{x}) = \int_{\text{surface}}  \mathbf{E}_d(\mathbf{r}'; \mathbf{p}) ~\exp{\left(-i \, \mathbf{k}_{DM} \mathbf{r} \right)} ~ d \mathbf{r} 
\end{equation}
and the magnetic field analogous, while $\mathbf{p}$ is chosen as above (and thus depends on $\mathbf{r}$), and $\mathbf{r}'~=~ \mathbf{x}-\mathbf{r}$. 
The factor $\exp{\left(-i \, \mathbf{k}_{DM} \mathbf{r} \right)}$ accounts for the spatial phase differences between the elementary waves induced by the dark matter wave. 
The intensity on the detector can be then obtained by evaluating the time-averaged Poynting-vector $\langle\mathbf{S}\rangle \cdot \hat{\mathbf{n}}_{det}$ while
\begin{equation}
\label{eq:poynting-vector}
\langle\mathbf{S}\rangle = \text{Re}(\mathbf{E} \times \mathbf{B}^*).
\end{equation}

\section{Stationary Case}
\begin{figure}
	\centering
	\centerline{\includegraphics[width=\textwidth]{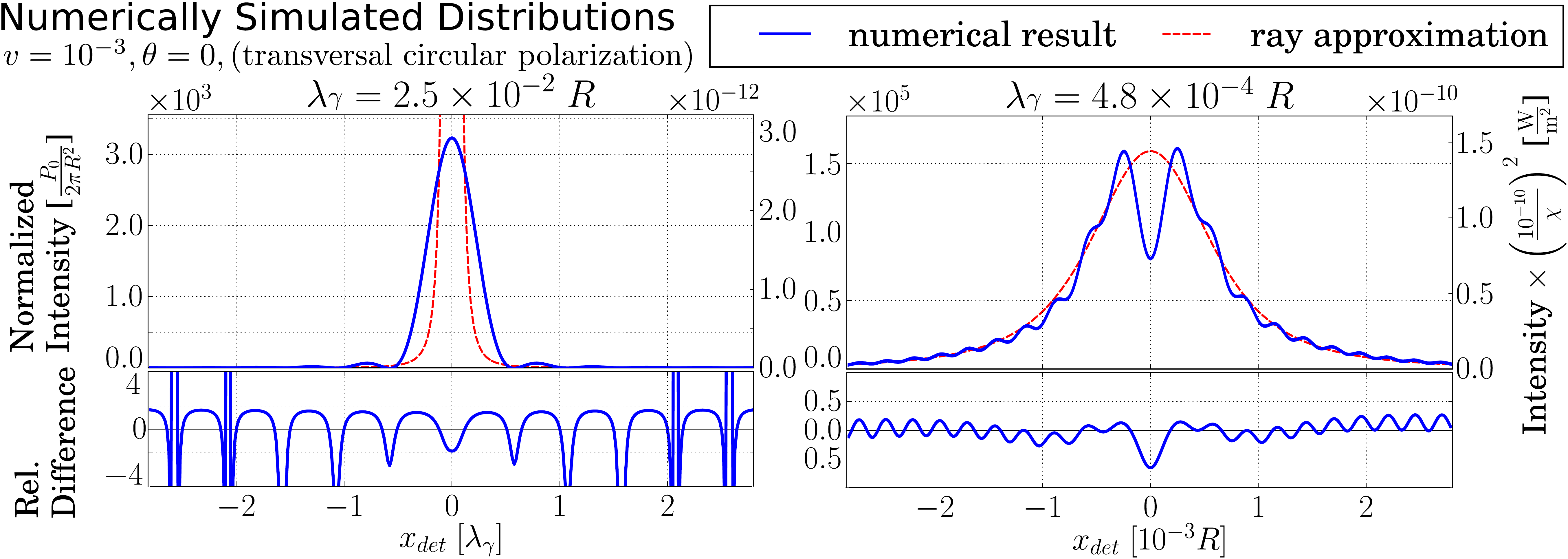}}
	\caption{Exemplary numerical results. Left: Stationary Distribution for high photon wavelength, Right: Distribution for wavelength comparable to distribution width.}
	\label{fig:distributions}
\end{figure}
\label{sec:stationary}
When the dark matter wavelength exceeds the setup size, i.e. $k_{DM}~R=m_X v~R\ll1$, dark matter oscillations appear stationary in the setup and directional resolution will be lost. 

If we further incoherently average over all dark matter polarizations and consider a spherical cap (as typical in most experimental setups), the direction of the field vectors becomes unimportant and we may use a simplified scalar ansatz
\begin{equation}
\label{eq:theo-scalar-integral}
I(\mathbf{x}) = \left|\int_{\text{surface}} \Psi(\mathbf{r}') ~ \exp{\left(-i \, \mathbf{k}_{DM} \mathbf{r}   \right)} ~ d^2 \mathbf{r}\right|^2
~~~ ; ~~~
\Psi(\mathbf{r}') = \Psi_0 ~ k_{\gamma} ~ {\frac{\exp(i k_{\gamma} r' )}{r'}}.
\end{equation}

We can now easily compute the intensity in the center of the detector $I(0)$, since $\exp{\left(-i \, \mathbf{k}_{DM} \mathbf{r}   \right)}$ is just a constant phase and $r'=R$ constant over the whole sphere. We find
\begin{equation}
	I(0) \propto A_{dish} ~ (1 - \cos(\vartheta_{max})) ~ {k_\gamma}^2 ~~ ; ~~{\rm FWHM} \propto \frac{\lambda_\gamma}{\sqrt{1 - \cos(\vartheta_{max})} },
\end{equation}
while the width dependence is obtained recalling that the total emitted power is $\propto A_{dish}$.

\section{Transitional Regime}
To infer effects for arbitrary hidden photon mass, we evaluated Eq.~\eqref{eq:poynting-vector} numerically. Our results are consistent with ray approximation for small wavelengths and Sec.~\ref{sec:stationary} for large wavelengths.

\begin{wrapfigure}[14]{R}{0.48\textwidth}
	\vspace{-1em}
	\centerline{\includegraphics[width=0.48\textwidth]{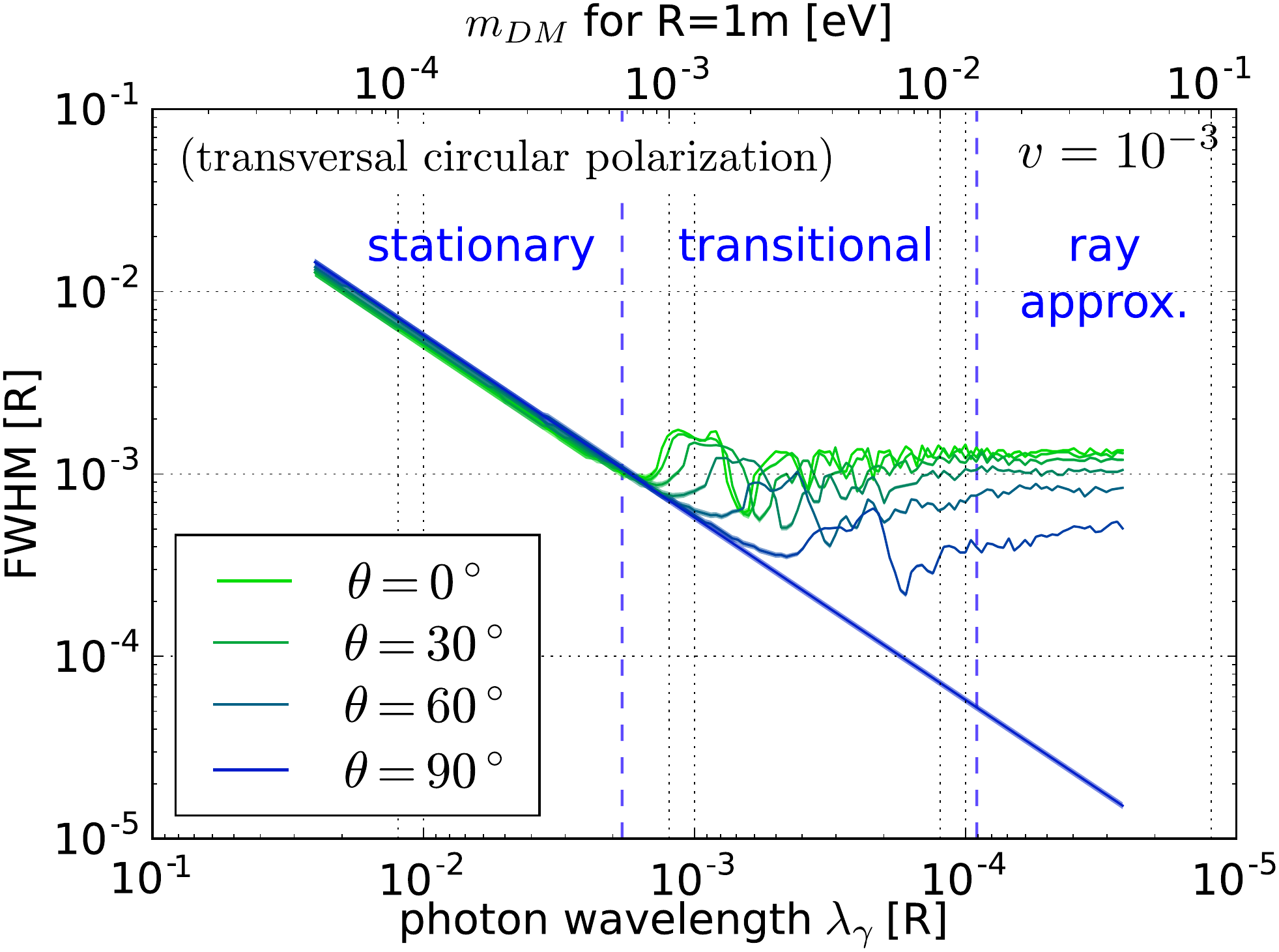}}
	\caption{Transition of width between stationary distribution and ray approximation for different incident angles and a full half sphere.}
	\label{fig:transition-width}
\end{wrapfigure}
For intermediate wavelengths smaller than expected distribution widths, the results from ray approximation get reproduced with overlaying interference effects as exemplified on the right side of Fig.~\ref{fig:distributions}.

We evaluated major distribution properties for a broad range of wavelengths, such as the distribution full width at half maximum (FWHM) shown in Fig.~\ref{fig:transition-width}. In the stationary regime the width is consistently proportional to the photon wavelength and converges for smaller wavelengths towards the ray approximated width in an oscillatory manner due to interference effects. 

\section{Conclusion} 
All numerical results confirm the stationary and ray approximated widths 
\begin{equation}
	{\rm FWHM}_{\rm sta.} \sim \frac{\lambda_\gamma}{\sqrt{1 - \cos(\vartheta_{sp,max})}} ~~~~; ~~~~ {\rm FWHM}_{\rm ray} \sim  \max(\tan(\vartheta_{sp,max}); 1) R v.
\end{equation}
Comparing them also determines which of the two is dominant. The transition between stationary case and ray approximation happens from where both quantities are roughly the same to where the stationary width becomes one order smaller than the ray approximated width.
Specifying the setup geometry by
\begin{equation}
	\hat{R} := \sqrt{1 - \cos(\vartheta_{sp,max})} \max(\tan(\vartheta_{sp,max}); 1) R
\end{equation}
we find that the expected distributions may be classified as
\begin{equation}
	 \bigg\{ \begin{array}{lcl}
		\hat{R} ~ \lesssim ~ 0.5 \lambda_\gamma / v = 1 \times \pi / (v~m_X) & \text{stationary}
		\\
		\hat{R} ~ \gtrsim ~ 10 \lambda_\gamma / v = 20 \times \pi / (v~m_X) & \text{ray approximated}
		\end{array}
\end{equation}
For experiments aiming for directional sensitivity, $v$ should be set to the smallest velocity one aims to be sensitive for, while for discovery experiments it is sufficient to choose $v$ as roughly the maximum dark matter velocity
in order to infer the expected distribution width.

\newpage
	


\end{document}